\documentclass[journal]{IEEEtran}

\ifCLASSINFOpdf
\else
   \usepackage[dvips]{graphicx}
\fi
\usepackage{url}

\usepackage{graphicx}
\usepackage{amssymb}	
\usepackage{seqsplit}  
\usepackage{bm}			
\usepackage{amsmath}	
\usepackage{graphicx} 	
\usepackage{subfigure} 	
\usepackage{svg}		
\usepackage{epstopdf}	
\usepackage{balance}	
\usepackage{cite}		
\begin{document}

\title{Underdetermined DOA Estimation of Off-Grid Sources Based on the Generalized Double Pareto Prior}

\author{Yongfeng Huang, Zhendong Chen, Kun Ye,  \IEEEmembership{Graduate Student Member, IEEE}, Lang Zhou, and Haixin Sun, \IEEEmembership{Sensior Member, IEEE}

\thanks{This work was supported by the National Natural Science Foundation of China under Grants 62271426, and the Natural Science Foundation of Fujian Province under Grants No. 2020J01003. ($\emph{Corresponding author: Haixin Sun.}$)}

\thanks{	
	Yongfeng Huang, Zhendong Chen, Kun Ye, and Haixin Sun are with the School of Informatics, Xiamen University, Xiamen 361005, China, and also with the Key Laboratory of Southeast Coast Marine Information Intelligent Perception and Application, Ministry of Natural Resources, Xiamen 361005, China(e-mail:huangyongfeng@stu.xmu.edu.cn; chenzhendong@stu.xmu.edu.cn; yekun@stu.xmu.edu.cn; hxsun@xmu.edu.cn).}

\thanks{
	Lang Zhou is with the School of Electronic Science and Engineering, Xiamen University, Xiamen 361005, China(e-mail: zhoulang@stu.xmu.edu.cn).}
}

\markboth{IEEE SIGNAL PROCESSING LETTERS, Vol. XX, No. XX, XXXX}
{Shell \MakeLowercase{\textit{et al.}}: Bare Demo of IEEEtran.cls for IEEE Journals}
\maketitle

\begin{abstract} 
In this letter, we investigate a new generalized double Pareto based on off-grid sparse Bayesian learning (GDP-OGSBL) approach to improve the performance of direction of arrival (DOA) estimation in underdetermined scenarios. The method aims to enhance the sparsity of source signal by utilizing the generalized double Pareto (GDP) prior. Firstly, we employ a first-order linear Taylor expansion to model the real array manifold matrix, and Bayesian inference is utilized to calculate the off-grid error, which mitigates the grid dictionary mismatch problem in underdetermined scenarios. Secondly, an innovative grid refinement method is introduced, treating grid points as iterative parameters to minimize the modeling error between the source and grid points. The numerical simulation results verify the superiority of the proposed strategy, especially when dealing with a coarse grid and few snapshots.
\end{abstract}


\IEEEpeerreviewmaketitle

\section{Introduction} 
\IEEEPARstart{D}{irection} of arrival (DOA) with sparse arrays is an important research in signal processing and is applicable in underwater sonar, telecommunications, radar, and human-computer interaction systems \cite{Zhou_EDNA, Zhou_RENA, Ye_IET}. For DOA estimation using sparse arrays, the spatial smoothing MUSIC (SS-MUSIC) is implemented for super-resolution spectral estimation in the coarray domain \cite{Remark_SS_MUSIC, SS_MUSIC}. However, subspace algorithms such as MUSIC \cite{MUSIC} and ESPRIT \cite{ESPRIT} count on adequate snapshots and high signal-to-noise ratio (SNR) to achieve accurate performance. Subsequently, sparse signal recovery (SSR) methods overcome the shortcomings of subspace methods and are widely used, which exhibits a number of excellent merits compared to traditional subspace methods, such as flexibility in leveraging the potential structure of the signal for recovery, no need for noise statistics, and improved robustness to noise \cite{Outlier, an_efficient}.

Sparse Bayesian learning (SBL) is one of the most attractive SSR methods \cite{Tipping, FVBI}. The SBL problem involves maximum a posteriori estimation with sparsity information obtained from a sparse prior for the hidden variable \cite{18_OGSBL_NA}, where Laplace and Gaussian priors are two popular choices for enforcing sparsity. The Laplace prior is deemed more effective to promote the sparsity of solution compared to  the Gaussian prior, while the Gaussian prior offers a simpler update rule and fewer parameter adjustments \cite{Bayesian_Laplace}. Wang $et \ al.$ \cite{CGDP_SBL1} introduced a generalized double Pareto (GDP) prior with sharper peaks at the origin of the probability density function (PDF), yielding a heavier sparsity than the traditional distribution. However, DOA estimation accuracy remains low when using coarse grids \cite{Sinc_Interpolation_TVT}.

In most scenarios, One issue with the SBL-based approach is the off-grid problem. Actual sources may not fall precisely on the predefined discrete grids, which impairs the performance and recognizability of the SBL algorithm. In \cite{off-grid-Gaussian-distribution}, the off-grid error is assumed to be a perturbation added to the measurement matrix and obeys a Gaussian distribution. However, the off-grid error distribution is closer to a uniform distribution rather than a Gaussian distribution in most practical applications \cite{rootSBL, OGSBL}. Dai $et \ al.$ \cite{rootSBL} update each grid point using the polynomial root to eliminate the off-grid error. In \cite{OGSBL}, the authors proposed a method called off-grid SBL (OGSBL). By exploiting the sparse information of the interest signal from a Bayesian perspective, the SBL method outperforms the ${\ell}_1$-norm method. However, a major problem is that its performance depends significantly on the trade-off between accuracy and computational complexity.

This letter introduces a novel method based on GDP prior, which uses a three-stage sparse prior model to enhance the spatial sparsity of the interest signals and optimizes the grid iteratively using the expectation-maximization (EM) method. Additionally, we propose a new grid refinement method to address the modeling error between the real source and the nearest grid point. Simulation experiments validate a significant improvement in DOA estimation performance using the proposed method, with a further reduction in the modeling error compared to the existing algorithm.

\section{Probelm Formulation}
Consider $K$ narrowband far-field signals $\mathbf{x}_k(t),k=1,\ldots,K$ impinging on an $M$-element sparse linear array from $\theta_k,k=1,\ldots,K$. These signals are presumed to have zero mean, exhibit temporal whiteness, and originate from spatially uncorrelated sources. $a_md$ denotes the $m$th sensor position, where $d$ is the unit inter-element spacing. The set of all sensors can be denoted as:
\begin{equation}
	\label{eq1}
	\mathbb{S}=\{{a}_{m}d, \ 0\leq {m} \leq {M-1}\}.
\end{equation}

To identify potential source candidates, the range of interest for DOA is divided evenly into $N$ grids, with $\mathbf{\Theta}=\left[\tilde{\theta}_1, \ldots, \tilde{\theta}_N \right]$ being the predefined range. The power of each grid is then estimated. The true DOA will fall within $\mathbf{\Theta}$ when the grids are dense enough. Subsequently, we construct an overcomplete steering matrix with $N \gg K > M$:
\begin{equation}
	\label{on-grid-model}
	\mathbf{Y}=\mathbf{A}_{\mathbf{\Theta}}{\mathbf{X}}+\mathbf{N},
\end{equation}
where $\mathbf{Y} \triangleq \left[ \mathbf{y}(1),\ldots ,\mathbf{y}(L) \right]$, $\mathbf{N} \triangleq \left[ \mathbf{n}(1),\ldots ,\mathbf{n}(L) \right]$, $\mathbf{{X}} \triangleq \left[ \mathbf{x}(1),\ldots ,\mathbf{x}(L) \right]$, $\mathbf{A}_{\mathbf{\Theta}}=\left[\mathbf{a}\big({\tilde{\theta}_1}\big), \ldots, \mathbf{a}\big({\tilde{\theta}_N}\big)\right]$, $\mathbf{Y}, \mathbf{N} \in {\mathbb{C}}^{M\times L}$, $\mathbf{X} \in {\mathbb{C}}^{N\times L}$, $\mathbf{A}_{\mathbf{\Theta}} \in \mathbb{C}^{M \times N}$, $L$ denotes the number of snapshots. $\mathbf{y}(t)$, $\mathbf{x}(t)$, and $\mathbf{n}(t)$ represent the array output vector, signal vector, and noise vector at the time $t$, respectively. $\mathbf{a}\left( {{{\tilde{\theta }}}_{n}} \right)={{\left[ {{e}^{-\frac{2\pi j{{a}_{0}}d}{\lambda }\sin \left( {{{\tilde{\theta }}}_{n}} \right)}},\ldots ,{{e}^{-\frac{2\pi j{{a}_{M-1}}d}{\lambda }\sin \left( {{{\tilde{\theta }}}_{n}} \right)}} \right]}^{T}}$ is the steering vector impinging from ${{\tilde{\theta }}_{n}}$ with the wavelength of signals $\lambda$. However, the true DOA commonly does not fall on the discrete grid, which leads to the dictionary mismatch problem. 

To address this issue, we approximate the steering vector of the impinging angle from ${\theta}_k$ by using the first-order Taylor expansion of the nearest grid point $\theta_{n_{k}}$, e.g., $\mathbf{a}({\theta}_k) \approx \mathbf{a}(\tilde{\theta}_{n_k})+\mathbf{a}^{'}(\tilde{\theta}_{n_k}){({\theta}_k-\tilde{\theta}_{n_k})}$. $\mathbf{{a}'}(\theta)$ represents the first-order derivative of $\theta$. This approximation allows for a more accurate representation of the received signal. The model (\ref{on-grid-model}) can then be formulated as:
\begin{equation}
	\label{off-grid-model}
	\mathbf{Y}=\bm{\Phi}(\bm{\beta}){\mathbf{X}}+\mathbf{N},
\end{equation}
where $\bm{\Phi}(\bm{\beta})=\mathbf{A}_{\mathbf{\Theta}}+\mathbf{B}diag\{\bm{\beta}\}$, $\bm{\Phi \left( {\beta} \right)}, \mathbf{A}_{\mathbf{\Theta}}, \mathbf{B} \in \mathbb{C}^{M \times N}$. $\mathbf{B}=\left[\mathbf{a}^{'}(\tilde{\theta}_{1}), \ldots, \mathbf{a}^{'}(\tilde{\theta}_{N})\right]$, $\bm{\beta}=\left[{{\beta}_1, \ldots, {\beta}_N}\right]^T$. To aid the clarity, the following text will refer $\bm{\Phi \left( {\beta} \right)}$ as $\bm{\Phi}$. $\bm{\beta}$ is a vector whose elements are all zero except for $\beta_{n_k}=\theta_k-\tilde{\theta}_{n_k}$ in the $n_k$th element with $\beta_{n_k} \in \left[-\frac{r}{2}, \frac{r}{2} \right]$. $r$ is the step size of the search grid.

\section{Proposed Method}
\subsection{Three-stage Hierachical Prior Model}	

Firstly, we introduce a three-layer prior model of GDP prior as the prior distribution for the interest signal $\mathbf{X}$. GDP prior distribution favors a sparser solution than Laplace distribution and student's $t$-distribution \cite{student-t-distribution}. The elements of $\mathbf{X}$ are assumed to follow an independent complex Gaussian distribution \cite{TIM-GDP}:
\begin{equation}
	\label{X-distribution}
	p\left( \mathbf{X}|\mathbf{\Lambda } \right)=\prod\limits_{t=1}^{L}{\prod\limits_{n=1}^{N}{\mathcal{C}\mathcal{N}\left( {{\mathbf{X}}_{n,t}}|0,{{\delta }_{n}} \right)}},
\end{equation}
where $\bm{\Lambda}=diag\{\bm{\delta}\}$, $\bm{\delta}={{\left[ {{\delta }_{1}},{{\delta }_{2}},\ldots ,{{\delta }_{N}} \right]}^{T}}$ is the variance of each row elements in $\mathbf{X}$. $n$ and $t$ are the row and column index of $\mathbf{X}$. $diag\left\{ \cdot  \right\}$ denotes an operator of diagonal matrix. Then, the hyperparameters $\delta_n, n=1, \ldots, N$ follow an independent Gamma distribution:
\begin{equation}
	\label{eq5}
	p\left( \bm{\delta }|\bm{\eta } \right)=\prod\limits_{n=1}^{N}{\Gamma \left( {{\delta }_{n}}|\frac{3}{2},\frac{\eta _{n}^{2}}{4} \right)},
\end{equation}
with $\Gamma \left( x|a,b \right)=\left( \frac{{{b}^{a}}}{\Gamma \left( a \right)} \right){{x}^{a-1}}\exp \left( -bx \right)$, $\Gamma \left( a \right)=\int_{0}^{\infty }{{{x}^{a-1}}\exp \left( -x \right)dx}$ and $\bm{\eta }={{\left[ {{\eta }_{1}}, \ldots, {{\eta }_{N}} \right]}^{T}}$. Similarly, each elements in $\bm{\eta}$ also follow an independent Gamma distribution:
\begin{equation}
	\label{eq6}
	p\left( \bm{\eta } \right)=\prod\limits_{n=1}^{N}{\Gamma \left( \sigma ,\sigma  \right)},
\end{equation}
where $\sigma$ is a positive constant that controls the sparsity of $\mathbf{X}$. In result, the marginal probability function of the sparse signal $\mathbf{X}$ can be written as \cite{CGDP_SBL1}:
\begin{equation}
	\label{eq7}
	\begin{split}
		p\left( {{\mathbf{X}}_{n,t}};\sigma \right) 
		&=\prod\limits_{n=1}^{N}{\iint{p\left( \mathbf{X}|\bm{\Lambda } \right)p\left( \bm{\delta }|\bm{\eta } \right)p\left( \bm{\eta } \right)d}}{{\delta }_{n}}d{{\eta }_{n}} \\ 
		&=\frac{1}{2\pi }\prod\limits_{n=1}^{N}{\frac{\left( \sigma +1 \right)/\sigma }{{{\left( \left| {{\mathbf{X}}_{n,t}} \right|/\sigma +1 \right)}^{\left( \sigma +1 \right)+1}}}}.
	\end{split}
\end{equation}

\subsection{Bayesian Inference}						

By combining model (\ref{off-grid-model}) with the assumption of circular symmetric complex Gaussian noises, there is $p\left( \mathbf{N} \right)=\prod\limits_{m=1}^{M}{\prod\limits_{t=1}^{L}{\mathcal{C}\mathcal{N}\left( {{\mathbf{N}}_{m,t}}|0,{{\alpha }^{-1}} \right)}}$ with the noise precision $\alpha$. The distribution for the off-grid error $\beta_{n_k}$ is assumed to be uniformly distributed between $\left[-\frac{r}{2}, \frac{r}{2}\right]$. Finally, we have:
\begin{equation}
	\label{Y-likelihood-probability}
	p\left( \mathbf{Y}|\mathbf{X},\alpha ,\bm{\beta } \right)=\prod\limits_{t=1}^{L}{\mathcal{C}\mathcal{N}}\left( \mathbf{y}(t)|\mathbf{\Phi }\mathbf{x}(t),{{\alpha }^{-1}}{{\mathbf{I}}_{M}} \right),
\end{equation}
where $I_M$ denotes an unit array of dimension $M$. The most likely value of $\bm{\delta },\bm{\eta },\alpha ,\bm{\beta }$ should be equivalent to the value of hyperparameters that maximizes the posterior probability $p\left( \mathbf{X},\bm{\delta },\bm{\eta },\alpha ,\bm{\beta }|\mathbf{Y} \right)$, but it cannot be computed explicitly. Therefore, we utilize EM algorithm to perform Bayesian inference.

In the E-step, the sparse signal $\mathbf{X}$ is treated as a hidden variable. The posterior probability of $\mathbf{X}$ obeys a multivariate complex Gaussian distribution according to (\ref{X-distribution})-(\ref{Y-likelihood-probability}):
\begin{equation}
	\label{eq9}
	p\left( \mathbf{X}|\mathbf{Y},\bm{\delta },\bm{\eta },\alpha ,\bm{\beta } \right)=\prod\limits_{t=1}^{L}{\mathcal{C}\mathcal{N}}\left( \bm{\mu },\mathbf{\Sigma } \right),
\end{equation}
where $\bm{\mu }=\bm{\Lambda }{{\bm{\Phi }}^{H}}\bm{\Sigma }_{\mathbf{Y}}^{-1}\mathbf{Y}$, $\mathbf{\Sigma }=\mathbf{\Lambda }-\mathbf{\Lambda }{{\mathbf{\Phi }}^{H}}\mathbf{\Sigma }_{\mathbf{Y}}^{-1}\mathbf{\Phi \Lambda }$ and ${{\mathbf{\Sigma }}_{\mathbf{Y}}}={{\alpha }^{-1}}{{\mathbf{I}}_{M}}+\mathbf{\Phi \Lambda }{{\mathbf{\Phi }}^{H}}$. The loss function is expressed as the logarithm of the joint probability density function, with its expectation taken:
\begin{equation}
	\label{joint-probability}
	\begin{split}
		Q(\bm{\delta}, &\bm{\eta }, \alpha, \bm{\beta}) \\
		& = {\left< \mathrm{ln}{\left[
				p(\mathbf{Y},\mathbf{X}, \bm{\delta},\bm{\eta},\alpha ,\bm{\beta})\right]}\right>}_{p( \mathbf{X}|\mathbf{Y}, \bm{\delta}, \bm{\eta}, \alpha, \bm{\beta})} \\
		& \propto {\left< \mathrm{ln}{\left[
				p(\mathbf{Y}|\mathbf{X}, \alpha, \bm{\beta})p\left( \mathbf{X}|\bm{\delta} \right)
				p\left(\bm{\delta}|\bm{\eta} \right)p\left(\bm{\eta}\right)\right]}\right>}_{p( \mathbf{X}|\mathbf{Y}, \bm{\delta}, \bm{\eta}, \alpha, \bm{\beta})},
	\end{split}
\end{equation}
where $\langle\cdot\rangle_p$ denotes an expectation operator with respect to the probability $p$. 

In M-step, we obtain the partial derivative of ${\delta}_n$ for (\ref{joint-probability}):
\begin{equation}
	\label{delta-derivative}
	\begin{split}
		\frac{\partial Q\left( \bm{\delta }, \bm{\eta}, \alpha, \bm{\beta} \right)}{\partial {{\delta}_{n}}}=-\frac{L}{{{\delta }_{n}}}+\frac{{{\bm{\mu}}_{n\cdot }}\bm{\mu}_{n\cdot }^{H}+L{{\bm{\Sigma}}_{nn}}}{\delta _{n}^{2}}-\frac{\eta _{n}^{2}}{4}+\frac{1}{2{{\delta }_{n}}},
	\end{split}
\end{equation}
Where ${{\bm{\Sigma}}_{nn}}$ represents the $n$th element on the diagonal of $\bm{\Sigma}$ and ${{\bm{\mu}}_{n\cdot}}$ represents the $n$th element of the row of $\bm{\mu}$. 

The fixed-point strategy is employed to boost convergence speed. Let $W_n=-1+{\frac{\bm{\Sigma}_{nn}}{\delta_n}}$, then bringing it into (\ref{delta-derivative}) yields $\frac{\partial Q\left( \bm{\delta }, \bm{\eta}, \alpha, \bm{\beta} \right)}{\partial {{\delta}_{n}}}=\frac{L{{W}_{n}}}{{{\delta }_{n}}}+\frac{{{\bm{\mu }}_{n\cdot }}\bm{\mu }_{n\cdot }^{H}}{\delta _{n}^{2}}-\frac{\eta _{n}^{2}}{4}+\frac{1}{2{{\delta }_{n}}}$. Therefore, the fixed-point update expression for ${\delta}_{n}$ is:
\begin{equation}
	\label{eq12}
	\begin{split}
		\delta _{n}^{\left( new \right)}
		&=\eta _{n}^{-2}\left( 1+2L{{W}_{n}}+2\sqrt{{{\left( 1/2+L{{W}_{n}} \right)}^{2}}+{{\bm{\mu }}_{n\cdot }}\bm{\mu }_{n\cdot }^{H}\eta _{n}^{2}} \right).
	\end{split}
\end{equation}

For ${{\eta}_{n}}$ and $\alpha$, we can obtain the update functions for both by taking the zeros of the partial derivatives of (\ref{joint-probability}):
\begin{equation}
	\label{eq13}
	\begin{split}
		\eta _{n}^{\left( new \right)}
		&=\frac{-\sigma +\sqrt{{{\sigma }^{2}}+2{{\delta }_{n}}\left( \sigma +2 \right)}}{{{\delta }_{n}}},\\
		{{\alpha }^{\left( new \right)}}
		&=\frac{ML}{\left\| \mathbf{Y}-{\bm{\Phi}}\bm{\mu } \right\|_{F}^{2}+Ltr\left( \mathbf{D}-\mathbf{D\Sigma }_{\mathbf{Y}}^{-1}\mathbf{D} \right)},
	\end{split}
\end{equation}
where $tr\left( \cdot  \right)$ and ${{\left\| \cdot  \right\|}_{F}}$ denote the trace operator and Frobenius norm, respectively, with $\mathbf{D}=\mathbf{\Phi \Lambda }{{\mathbf{\Phi }}^{H}}$.

\subsection{New Off-Grid Refinement}				

To optimize the off-grid error ${\beta }_{{n}_{k}}$, the objective is to maximize $E\left\{ \ln p\left( \mathbf{Y}|\mathbf{X},{\alpha},\bm{\beta } \right)p\left( \bm{\beta } \right) \right\}$, thereby equivalent the maximization of $E\left\{ \ln p\left( \mathbf{Y}|\mathbf{X}, {\alpha}, \bm{\beta} \right) \right\}$:
\begin{equation}
	\label{eq14}
	\begin{split}
		Q(\bm{\beta})
		\triangleq & {{\left\langle \ln p\left( \mathbf{Y}|\mathbf{X},{\alpha},\bm{\beta } \right) \right\rangle }_{p( \mathbf{X}|\mathbf{Y},\bm{\delta }, \bm{\eta}, \alpha, \bm{\beta})}} \\ 
		= &-ML\ln \pi +L\ln {\alpha} \\
		& -\sum\limits_{t=1}^{L}{{\alpha}{{\left[ \mathbf{y}(t)-\mathbf{\Phi}\mathbf{x}(t) \right]}^{H}}\left[ \mathbf{y}(t)-\bm{\Phi}\mathbf{x}(t) \right]}.
	\end{split}
\end{equation}

Thus, the maximization problem is transformed into the minimization of $E\left\{ \left\| \mathbf{y}(t)-\bm{\Phi}\mathbf{x}(t) \right\|_{2}^{2} \right\}$, where $||\cdot||_2$ represents the ${\ell}_2$ norm:
\begin{equation}
	\label{beta-loss-function}
	E\left\{ \left\| \mathbf{y}(t)-\bm{\Phi }\mathbf{x}(t) \right\|_{2}^{2} \right\}={{\bm{\beta}}^{T}}\bm{P\beta }-2{{\mathbf{Q}}^{T}}\bm{\beta}+C,
\end{equation}
where the expression of $\mathbf{P}$ and $\mathbf{Q}$ is:
\begin{equation}
	\label{eq16}
	\begin{split}
		\mathbf{P} 
		=&\Re \left\{ {{\mathbf{B}}^{H}}\mathbf{B}\odot(\bm{\mu}{{\bm{\mu}}^{H}}+\mathbf{\Sigma}) \right\}, \\
		\mathbf{Q} 
		= &\Re \left\{ \frac{1}{L}\sum\limits_{t=1}^{L}{diag\left(\bm{\mu}(t){{\mathbf{B}}^{H}}\left( \mathbf{y}(t)-{{\mathbf{A}}_{\bm{\Theta}}}\bm{\mu}(t)\right)\right)} \right\} \\
		 & -\Re \left\{ diag\left( {{\mathbf{B}}^{H}}{{\mathbf{A}}_{{\mathbf{\Theta}}}}\mathbf{\Sigma} \right) \right\},
	\end{split}
\end{equation}
where $\mathbf{a} \odot \mathbf{b}$ denotes the Hadamard (element-wise) product of $\mathbf{a}$ and $\mathbf{b}$. $\Re \left\{ \cdot  \right\}$ take the real part of a complex variable. 

Therefore, we solve for the zero by taking the derivative of (\ref{beta-loss-function}). If $\mathbf{P}$ is reversible, then ${{\bm{\beta }}^{(new)}}={{\mathbf{P}}^{-1}}\mathbf{Q}$; otherwise, we need to solve iteratively for each element in $\bm{\beta}$ (e.g, $\hat{\beta }_{n}^{new}=\frac{{{Q}_{n}}-\left( {{\mathbf{P}}_{n}} \right)_{-n}^{T}{\bm{{{\beta }_{-n}}}}}{{{P}_{nn}}}$), where ${{\mathbf{u}}_{-n}}$ represents the vector $\mathbf{u}$ after removing the $n$th element of the vector. Additionally, we remap $\bm{\beta}$ to $\left[-\frac{r}{2}, \frac{r}{2}\right]$ when it is out of range.

Note that the original off-grid method employs a first-order linear approximation for the measurement matrix $\bm{\Phi}$. The final DOA estimate combines the nearest grid point to the true source and the off-grid error. Although the first-order linear approximation provides a better representation of the actual array manifold matrix, the distance between the real DOA and the nearest grid point remains unaltered during iterative updates. Consequently, the modeling error introduced by (\ref{on-grid-model}) persists. To tackle this issue, we treat the grid points as an adjustable parameters and update them directly:
\begin{equation}
	\label{seta-update-function}
	\tilde{\theta}_{n}^{\left( new \right)}={{\tilde{\theta}}_{n}}+{{\beta }_{n}}, n=1, \ldots, K.
\end{equation}

Similarly, update the dictionary matrix $\mathbf{A}_{{\bm{\Theta}}^{(new)}}$ and its first-order derivative matrix $\mathbf{B}_{{\mathbf{\Theta}}^{(new)}}$, where ${{\bm{\Theta}}^{(new)}}=\left[ \tilde{\theta}_{1}^{(new)}, \ldots, \tilde{\theta}_{N}^{(new)} \right]$ is the sampling grids updated by (\ref{seta-update-function}):

\begin{equation}
	\label{eq18}
	\begin{split}
		{{\mathbf{A}}_{{{{\bm{\Theta }}}^{(new)}}}}
		&=\left[ \mathbf{a}\left( \tilde{\theta}_{1}^{(new)}\right), \mathbf{a}\left( \tilde{\theta }_{2}^{(new)} \right), \ldots , \mathbf{a}\left( \tilde{\theta }_{N}^{(new)} \right) \right], \\
		{{\mathbf{B}}_{{{{\bm{\Theta }}}^{(new)}}}}
		&=-\frac{2\pi j\mathbf{S}}{\lambda }\cos\left({\bm{\Theta}}^{(new)}\right) \odot {{\mathbf{A}}_{{{{\bm{\Theta }}}^{(new)}}}},
	\end{split}
\end{equation}
where $\mathbf{S}$ denotes the column vector consisting of the array position set $\mathbb{S}$ in ascending order.

During the iteration process, we opt to update (\ref{seta-update-function}) iteratively rather than updating $\bm{\Phi}$ alone. The grids gradually converge to the correct orientation after multiple iterations, thereby diminishing the modeling error. If the newly grid points fall within $\left[ \frac{{{{\hat{\theta }}}_{i-1}}+{{{\hat{\theta }}}_{i}}}{2},\frac{{{{\hat{\theta }}}_{i}}+{{{\hat{\theta}}}_{i+1}}}{2} \right]$, grid refinement is updated; otherwise, no update is performed.

After the SBL algorithm converges, the maximum $K$ peak clusters of the power ${{\delta}_{n}}, n=1,\ldots ,N$ are the estimated DOA angles ${{\tilde{\theta }}_{k}}, k=1,\ldots ,K$. To further eliminate the DOA estimation error, we consider a direct mapping equation between the estimated DOA angle $\tilde{\theta}$ and the power $\tilde{\delta}$ \cite{CGDP_SBL1}:
\begin{equation}
	\label{eq19}
	\begin{split}
		L(\delta)
		=&-L\ln \left| {{\mathbf{\Sigma }}_{\mathbf{Y}}} \right|-\sum\nolimits_{t=1}^{L}{\mathbf{y}(t)\mathbf{\Sigma}_{\mathbf{Y}}^{-1}\mathbf{y}(t)} \\
		&+\sum\nolimits_{n=1}^{N}{\left( -\frac{\eta_{n}^{2}}{4}{{\delta }_{n}}+\ln \sqrt{{{\delta}_{n}}} \right)}.
	\end{split}
\end{equation}

Then we define ${{u}_{k}}={{\mathbf{a}}^{H}}( {{{\tilde{\theta}}}_{k}})\mathbf{\Sigma}_{\mathbf{Y}-k}^{-1}\mathbf{a}({{\tilde{\theta}}}_{k})$, ${v}_{k}=L{{\mathbf{a}}^{H}}({{\tilde{\theta}}}_{k})\mathbf{\Sigma}_{\mathbf{Y}-k}^{-1}{{\mathbf{R}}_{\mathbf{Y}\mathbf{Y}}}\mathbf{\Sigma}_{\mathbf{Y}-k}^{-1}\mathbf{a}({{{\tilde{\theta }}}_{k}})$, where ${{\mathbf{R}}_{\mathbf{Y}\mathbf{Y}}}=\mathbf{Y}{{\mathbf{Y}}^{H}}/L$, and we have:
\begin{equation}
	\label{seta-delta-direct-mapping-function}
	\begin{split}
		L({\tilde{\theta }}_{k}, {\tilde{\delta}}_{k})
		=&-L\ln \left| {{\mathbf{\Sigma}}_{\mathbf{Y}-k}}+{{{\tilde{\delta}}}_{n}}a( {{{\tilde{\theta}}}_{k}} ){{a}^{H}}( {{{\tilde{\theta}}}_{k}}) \right| \\
		& -\sum\nolimits_{t=1}^{L}{\left( \mathbf{y}(t){{( {{\mathbf{\Sigma }}_{\mathbf{Y}-k}}+{{{\tilde{\delta }}}_{n}}({{{\tilde{\theta }}}_{k}}){{a}^{H}}( {{{\tilde{\theta }}}_{k}}) )}^{-1}}\mathbf{y}(t) \right)} \\ 
		& -\frac{\tilde{\eta }_{n}^{2}}{4}{{{\tilde{\delta }}}_{n}}+\ln \sqrt{{{{\tilde{\delta }}}_{n}}} \\ 
		\approx & -L\ln( 1+{{{\tilde{\delta }}}_{k}}{{u}_{k}})+\frac{{{v}_{k}}{{{\tilde{\delta }}}_{k}}}{1+{{u}_{k}}{{{\tilde{\delta }}}_{k}}}-\frac{\tilde{\eta }_{k}^{2}}{4}{{{\tilde{\delta }}}_{k}},
	\end{split}
\end{equation}
with ${{\mathbf{\Sigma }}_{\mathbf{Y}-k}}={{\mathbf{\Sigma }}_{\mathbf{Y}}}-{{\tilde{\delta }}_{k}}\mathbf{a}({{{\tilde{\theta }}}_{k}}){{\mathbf{a}}^{H}}( {{{\tilde{\theta}}}_{k}})$. Derive ({\ref{seta-delta-direct-mapping-function}}) to find the zero. The updated expression for ${{\bar{\delta }}_{k}}$ can be obtained:
\begin{equation}
	\label{delta-bar}
	\begin{split}
		{{{\bar{\delta}}}_{k}}
		= & \frac{2}{{{u}_{k}}\tilde{\eta }_{k}^{2}}\sqrt{{{\left( L{{u}_{k}}+\tilde{\eta }_{k}^{2}/2 \right)}^{2}}-\tilde{\eta }_{k}^{2}\left( L{{u}_{k}}+\tilde{\eta }_{k}^{2}/4-{{v}_{k}} \right)} \\ 
		& -\frac{2}{{{u}_{k}}\tilde{\eta }_{k}^{2}}\left( L{{z}_{k}}+\tilde{\eta }_{k}^{2}/2 \right),k=1,\ldots ,K. 
	\end{split}
\end{equation}

${{\bar{\delta }}}_{k}$ is a function of $\tilde{\theta}_k$ because $u_k$ and $v_k$ are function of $\tilde{\theta}_k$, respectively. Therefore, The final loss function $L\left( {{{\tilde{\theta }}}_{k}},{{{\tilde{\delta }}}_{k}} \right)$ is equivalent to the loss function $L\left( {{{\tilde{\theta }}}_{k}} \right)$ relative to $\tilde{\theta}_k$ . The DOA search process can be accomplished in the $k$th peak cluster by scanning with a small step size. For instance, ${{\tilde{\delta }}_{n}}\left( {{{\tilde{\theta }}}_{k-1}} \right)>{{\tilde{\delta }}_{n}}\left( {{{\tilde{\theta }}}_{k+1}} \right)$, consider implementing (\ref{delta-bar}) with a small step size within $\left[ {{{\tilde{\theta }}}_{k-1}},{{{\tilde{\theta }}}_{k}} \right]$ and updating one by one to obtain ${{\bar{\delta }}}_{k}$; conversely, execute (\ref{delta-bar}) within $\left[ {{{\tilde{\theta }}}_{k}},{{{\tilde{\theta }}}_{k+1}} \right]$.

\section{Performance of Evaluation}			
In this section, we conduct simulations to compare our method with rootSBL \cite{rootSBL}, OGSBL \cite{OGSBL}, and CGDP-SBL1 \cite{CGDP_SBL1}. The root mean square error (RMSE) serves as an indicator of the algorithm effectiveness and is defined as:
\begin{equation}
	\label{eq22}
	 RMSE=\sqrt{\frac{1}{RK}\sum\limits_{r=1}^{R}{\sum\limits_{k=1}^{K}{{{\left( {{{\hat{\theta }}}_{r,k}}-{{\theta }_{k}} \right)}^{2}}}}}, 
\end{equation} 
 where $R$ is the number of Monte Carlo trials and ${{\hat{\theta }}_{r,k}}$ is the DOA estimate for the $k$th source in the $r$th Monte Carlo experiment. In all simulations, $R$ is set to 200.

\subsection{Simulation of Setup}	
We consider an underdetermined scenario in which 10 incoherent narrowband sources are estimated using a 9-element extended coprime array. In this case, the sources are uniformly distributed between $\left[ -{{60}^{\circ }},{{60}^{\circ }} \right]$, all with the same power. The range of interest for DOA is $\left[-\frac{\pi}{2}, \frac{\pi}{2} \right]$. Setting ${{N}_{1}}=3$ and ${{N}_{2}}=4$, a co-prime array of $2N_1+N_2-1=9$ sensors is considered with sensor position set $\mathbb{S}=\left\{ 0d,3d,4d,6d,8d,9d,12d,16d,20d \right\}$, where $d=\frac{\lambda }{2}$ is the minimum unit of array element spacing. For fairness in algorithm comparison, we uniformly set the maximum number of iterations to 1000 and allowable error to $\varepsilon ={{10}^{-3}}$. In our method, we set $\sigma =0.1$.

\subsection{DOA Estimation Performance} 	

In the first simulation, Fig.1 show the results of a single Monte Carlo DOA estimation under the experimental conditions set to L=200, SNR=20dB, and step=$1^{\circ}$. As evident from the results of the single experiment, all 10 sources (exceeding the number of physical array elements) are clearly resolved and the pseudo-peaks of our method are very small.

\begin{figure}
	\centering
	\subfigure{
		\includegraphics[scale=0.26]{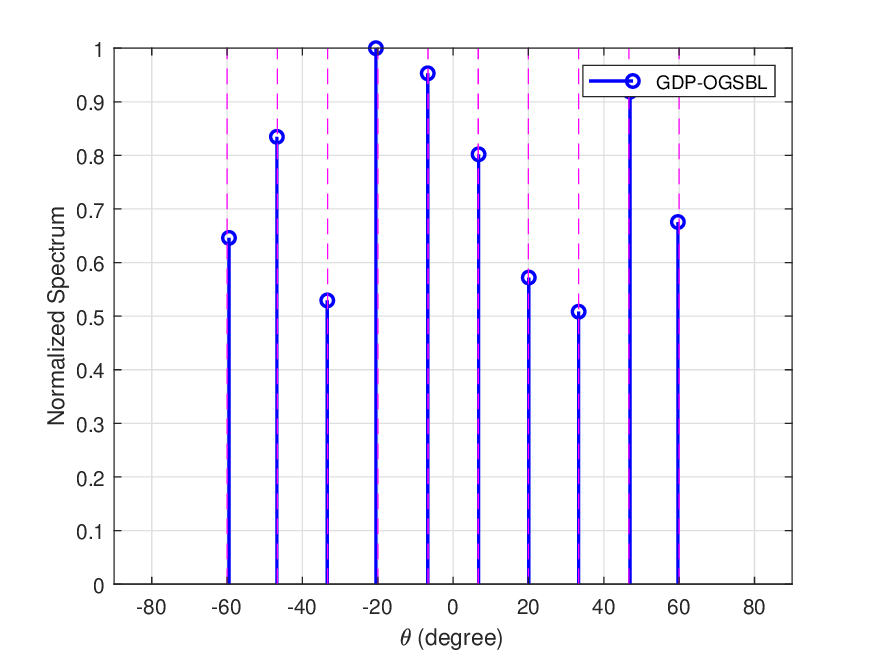} \label{fig1}
	}
	\quad
	\subfigure{
		\includegraphics[scale=0.26]{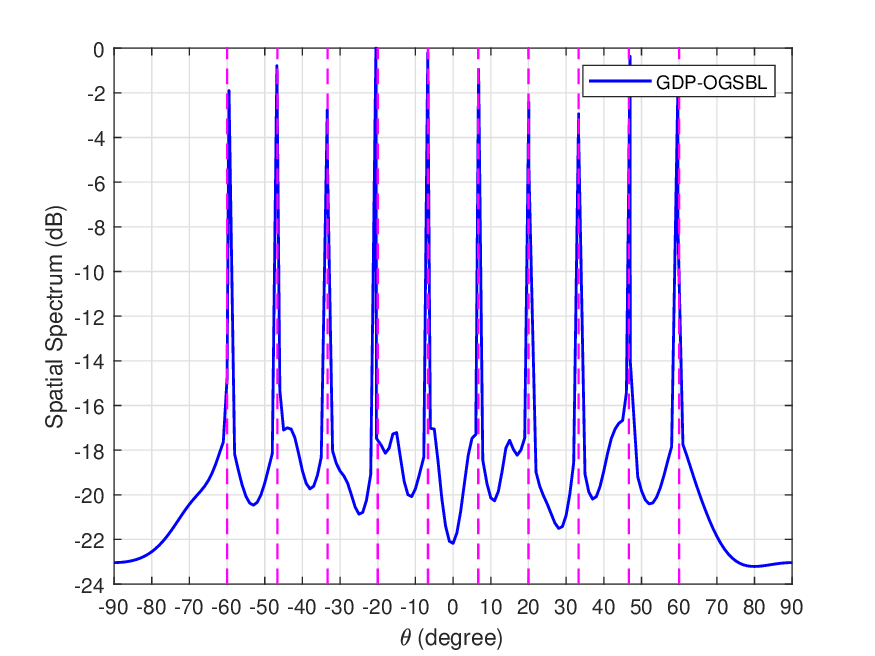} \label{fig2} 
	}
	\caption{Single DOA estimation result obtained by GDP-OGSBL, where the dotted lines represent the actual incident angles of the impinging signals. (a) Normalized power spectrum; (b) Spatial spectrum.
	}
\end{figure}

In the second simulation, we varied the SNR from -10 dB to 20 dB, with 200 snapshots and a grid step of $1^{\circ}$. To compared the RMSE versus different snapshots, the SNR is fixed at 0 dB, and the grid step is set to $1^{\circ}$, with the number of snapshots ranging from 50 to 500.

Fig.2 illustrates that the proposed method outperforms other algorithms in all cases. This is mainly attributed to the utilization of a generalized double Pareto prior, which promotes sparsity more than the Laplace prior. Additionally, the array manifold matrix's first-order Taylor expansion approximation is also introduced, which improves DOA estimation compared to the on-grid model.

\begin{figure}
	\centering
	\subfigure{
		\includegraphics[scale=0.26]{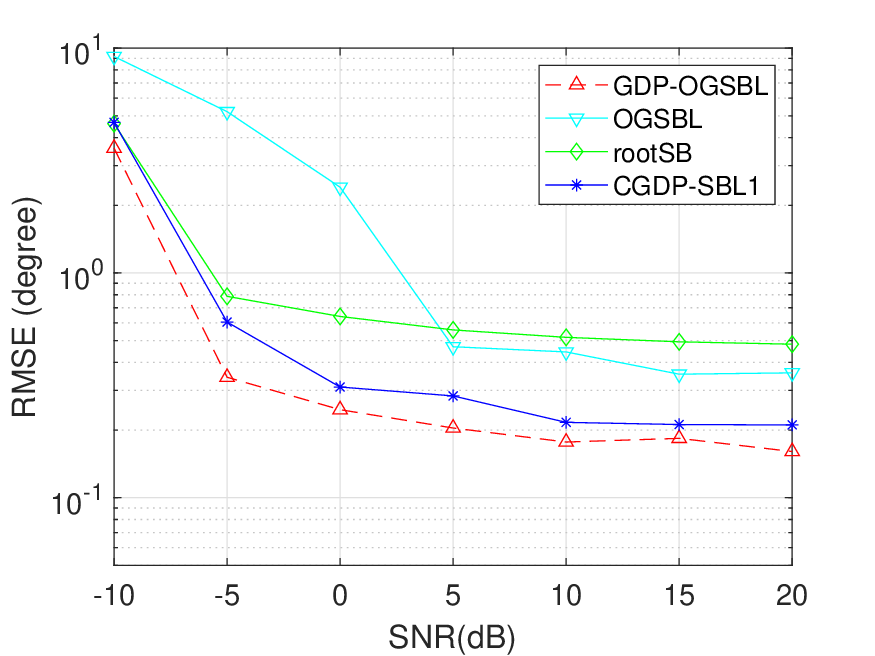} \label{fig3}
	}
	\quad
	\subfigure{
		\includegraphics[scale=0.26]{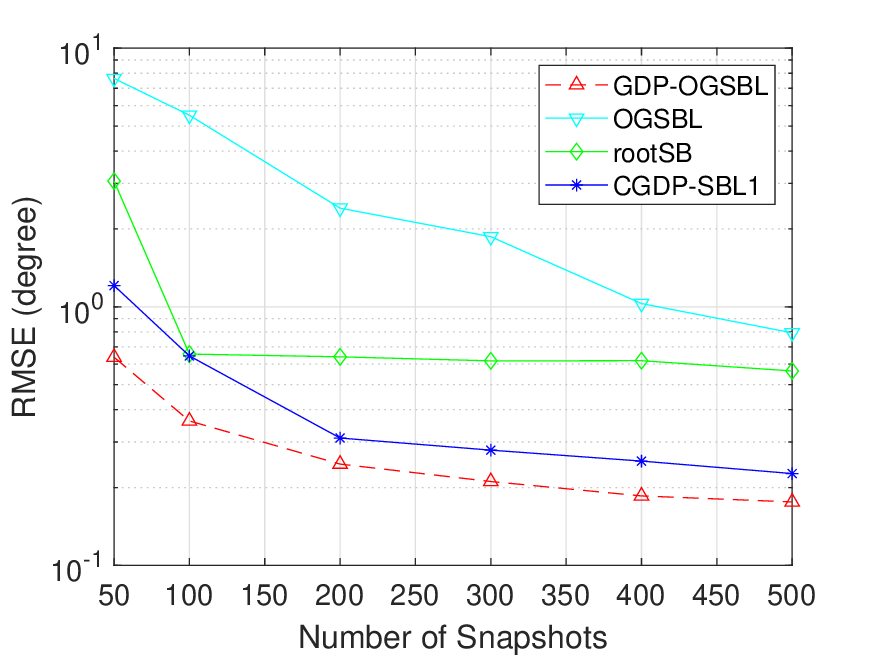} \label{fig4} 
	}
	\caption{Comparison of DOA estimation performance of different algorithms. (a) SNRs; (b) the number of snapshots.}
\end{figure}

The third simulation treats the grid step as a variable ranging from $1^\circ$ to $6^\circ$, with the number of snapshots fixed at 200 and the SNR fixed at 0 dB. In Fig.3, it is evident that the RMSE of all methods increases with the grid step. However, our method consistently maintains optimal performance, exhibiting minimal deterioration with the increase inthe grid step. Clearly, the iterative updating of grid points as variable parameters further reduces the modeling error. We suggest using coarser grids with a step size of $4^\circ$ to $6^\circ$ for quick DOA estimates without a significant decline in performance.

\begin{figure}
	\centering
		\includegraphics[scale=0.28]{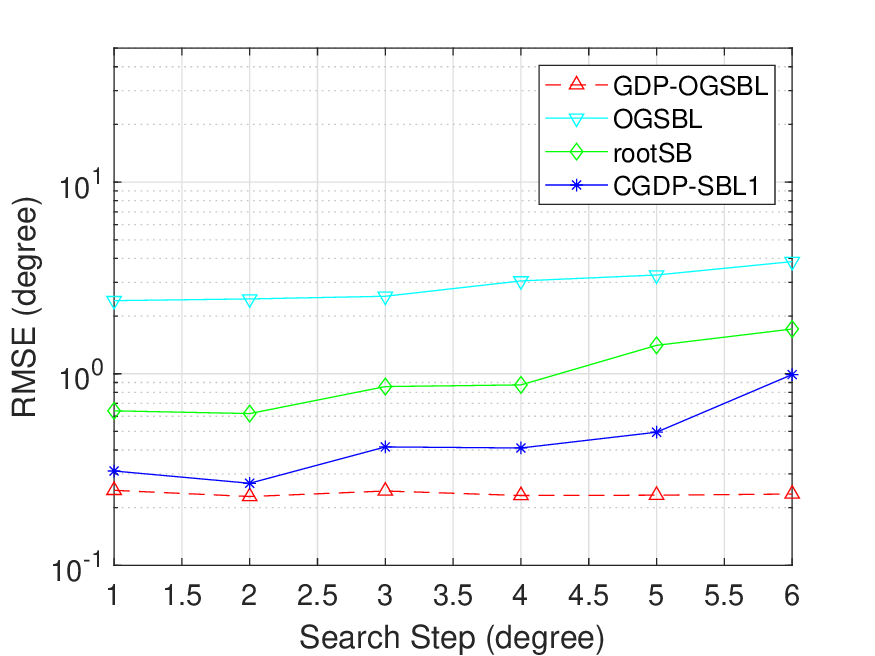}\label{search_step}
	\caption{Comparison chart of algorithms under different search step}
\end{figure}

\section{Conclusion}
To overcome the dictionary mismatch problem in underdetermined situations, we propose a sparse Bayesian learning method based on a GDP three-layer prior. This method utilizes a first-order Taylor expansion to alleviate the off-grid problem and continuously reduces the modeling error in each update iteration. Simulation results demonstrate a significant enhancement in DOA estimation performance compared to state-of-the-art algorithms, especially under coarse grids and few snapshots settings. In future work, DOA estimation in the 2-Dimensional wideband underdetermined case will be an interesting problem, and the group sparsity property of complex signals with different frequency bands will be taken into account.

\balance
\bibliographystyle{IEEEtran}
\bibliography{reference}
\end{document}